\documentstyle[galley,graphicx,epsf]{mn}
\newif\ifAMStwofonts
\ifoldfss
  \ifCUPmtlplainloaded \else
    \NewTextAlphabet{textbfit} {cmbxti10} {}
    \NewTextAlphabet{textbfss} {cmssbx10} {}
    \NewMathAlphabet{mathbfit} {cmbxti10} {} % for math mode
    \NewMathAlphabet{mathbfss} {cmssbx10} {} %  "   "    "
  \fi
  \ifAMStwofonts
    \ifCUPmtlplainloaded \else
      \NewSymbolFont{upmath} {eurm10}
      \NewSymbolFont{AMSa} {msam10}
      \NewMathSymbol{\upi}     {0}{upmath}{19}
      \NewMathSymbol{\umu}     {0}{upmath}{16}
      \NewMathSymbol{\upartial}{0}{upmath}{40}
      \NewMathSymbol{\leqslant}{3}{AMSa}{36}
      \NewMathSymbol{\geqslant}{3}{AMSa}{3E}

      \let\leq=\leqslant \let\le=\leqslant
       
    \fi
  \fi
\fi % End of OFSS

\ifnfssone
  \newmathalphabet{\mathit}
  \addtoversion{normal}{\mathit}{cmr}{m}{it}
  \addtoversion{bold}{\mathit}{cmr}{bx}{it}
  \newmathalphabet{\mathbfit} % math mode version of \textbfit{..}
  \addtoversion{normal}{\mathbfit}{cmr}{bx}{it}
  \addtoversion{bold}{\mathbfit}{cmr}{bx}{it}
  \newmathalphabet{\mathbfss} % math mode version of \textbfss{..}
  \addtoversion{normal}{\mathbfss}{cmss}{bx}{n}
  \addtoversion{bold}{\mathbfss}{cmss}{bx}{n}
  \ifAMStwofonts
    \ifCUPmtlplainloaded \else
      \UseAMStwoboldmath
      \makeatletter
      \new@mathgroup\upmath@group
      \define@mathgroup\mv@normal\upmath@group{eur}{m}{n}
      \define@mathgroup\mv@bold\upmath@group{eur}{b}{n}
      \edef\UPM{\hexnumber\upmath@group}
      \new@mathgroup\amsa@group
      \define@mathgroup\mv@normal\amsa@group{msa}{m}{n}
      \define@mathgroup\mv@bold\amsa@group{msa}{m}{n}
      \edef\AMSa{\hexnumber\amsa@group}
      \makeatother
      \mathchardef\upi="0\UPM19
      \mathchardef\umu="0\UPM16
      \mathchardef\upartial="0\UPM40
      \mathchardef\leqslant="3\AMSa36
      \mathchardef\geqslant="3\AMSa3E

      \let\leq=\leqslant \let\le=\leqslant

    \fi
  \fi
\fi % End of NFSS release 1

\ifnfsstwo
  \DeclareMathAlphabet{\mathbfit}{OT1}{cmr}{bx}{it}
  \SetMathAlphabet\mathbfit{bold}{OT1}{cmr}{bx}{it}
  \DeclareMathAlphabet{\mathbfss}{OT1}{cmss}{bx}{n}
  \SetMathAlphabet\mathbfss{bold}{OT1}{cmss}{bx}{n}
  \ifAMStwofonts
    \ifCUPmtlplainloaded \else
      \DeclareSymbolFont{UPM}{U}{eur}{m}{n}
      \SetSymbolFont{UPM}{bold}{U}{eur}{b}{n}
      \DeclareSymbolFont{AMSa}{U}{msa}{m}{n}
      \DeclareMathSymbol{\upi}{0}{UPM}{"19}
      \DeclareMathSymbol{\umu}{0}{UPM}{"16}
      \DeclareMathSymbol{\upartial}{0}{UPM}{"40}
      \DeclareMathSymbol{\leqslant}{3}{AMSa}{"36}
      \DeclareMathSymbol{\geqslant}{3}{AMSa}{"3E}

      \let\leq=\leqslant \let\le=\leqslant

    \fi
  \fi
\fi % End of NFSS release 2

\ifCUPmtlplainloaded \else
  \ifAMStwofonts \else % If no AMS fonts
    \def\upi{\pi}
    \def\umu{\mu}
    \def\upartial{\partial}
  \fi
\fi

\begin{document}
\title{Abundance analysis of cool extreme helium star: LSS\,3378}
\author[G. Pandey]
       {Gajendra Pandey and Bacham E. Reddy$^1$\\
       Indian Institute of Astrophysics, Bangalore 560034, India\\
       $^1$Visiting Observer, Cerro-Tololo Inter-American Observatory}
\date{Accepted .
      Received ;
      in original form 2006 }

\pagerange{\pageref{firstpage}--\pageref{lastpage}}
\pubyear{2006}

\maketitle

\label{firstpage}

\begin{abstract}

Abundance analysis of the cool extreme helium (EHe) star LSS\,3378 is presented.
The abundance analysis is done using LTE line formation and
LTE model atmospheres constructed for EHe stars.

The atmosphere of LSS\,3378 shows evidence of H-burning, 
He-burning, and $s$-process nucleosynthesis. The derived abundances
of iron-peak and $\alpha$-elements indicate absence of selective fractionation
or any other processes that can distort chemical composition of these
elements. Hence, the Fe abundance (log $\epsilon$(Fe) = 6.1)
is adopted as an initial metallicity indicator.
The measured abundances of LSS\,3378 are
compared with those of R Coronae Borealis (RCB) stars and with rest of the EHe stars as a group.

\end{abstract}


\section{Introduction}

The extreme helium stars (EHes) are supergiants with peculiar chemical
composition. The atmospheres of these supergiants with effective
temperature in the range 8000 -- 35000 K are devoid of hydrogen, and are
enriched with helium, carbon, and nitrogen with respect to the
atmospheres of normal main-sequence stars. Helium is the 
most abundant element in their atmospheres. There are about 21 known EHes.
There are 5 cool EHes, stars with effective temperatures 8000 to 13000 K, of
which 4 were analysed by Pandey et al. (2001). The remaining one cool EHe LSS\,3378
is analysed here.

Star No. 3378 (LSS\,3378) in the catalog of Stephenson and Sanduleak (1971)
was first identified as helium-rich B-type by Drilling (1973). Drilling
noticed the absence of Balmer absorption lines, presence of strong He\,{\sc i}
absorption lines, and the presence of C\,{\sc ii} line at 4267\AA\ in the
spectrum of LSS\,3378. Drilling also noted strong Mg\,{\sc ii} line at
4481\AA\ and Si\,{\sc ii} lines in absorption indicating a spectral class of
about B8 but, however, suggested somewhat earlier class than this because of the presence
of several weak O\,{\sc ii} lines.
Jeffery et al. (2001) reported the effective
temperature of LSS\,3378 to be about 10500 K by fixing $E_{B-V}$
using $IUE$ data and $UBV$ photometry. However, Jeffery et al. also reported that
the $IUE$ data for this star are very noisy.
Drilling et al. (1984) were the first to derive 
an effective temperature for LSS\,3378, obtaining 9400$\pm$500K, using the same 
procedure adopted by Jeffery et al.

R Coronae Borealis (RCB) stars, which are hydrogen-deficient F- and G-type supergiants, 
overlap in their effective temperatures with cooler EHes. With this abundance
analysis of LSS\,3378, we have the chemical composition of all the 5 known
cool EHes which are closely related to RCB stars. Our abundance analysis is based on
the high resolution optical spectrum.

\section{Observations}

The spectrum of LSS\,3378 was obtained on 2002 June 19 with the 4-m Blanco telescope
and the Cassegrain echelle spectrograph at CTIO in Chile. Three exposures of
25 minutes each, spectra covering the wavelength interval 5000 -- 8200\AA\ without gaps,
were recorded at a resolving power of R = 30,000.
The Image Reduction and Analysis Facility (IRAF) software packages were used
to reduce the recorded spectra.
The Th-Ar hollow cathode lamp provided lines for wavelength calibration.
A final spectrum was obtained by co-adding these three individual wavelength calibrated 
spectra from the three exposures. The maximum signal-to-noise (S/N) in the 
continuum (per pixel) of the co-added spectrum is between 200 and 250 at the 
middle of each echelle order.

The Na\,D lines in the spectrum of LSS 3378 are very strong and appear to
be saturated as seen in the spectra of other cool EHes: FQ\,Aqr, LS IV $-14^\circ109$,
BD $-1^\circ3438$, and LS IV $-1^\circ2$. These saturated Na\,D lines in the spectra of cool
EHes are certainly interstellar in origin. The Na\,D lines in the spectra of
cool EHes are stronger than in any other spectrum of normal star. Note that,
lines of ionized metals of the iron group are plentiful in the cool EHe's spectrum.
These lines are much stronger when compared with those observed in early A-type
and late B-type normal stars. This notable feature of the spectra of cool EHes
is attributable to the lower opacity in the atmosphere due to hydrogen deficiency.

\section{Abundance analysis - method}

\subsection{Procedure}

For the abundance analysis of LSS\,3378, same procedure is followed as
described in Pandey et al. (2001, 2004, 2006). The analysis 
uses line-blanketed hydrogen-deficient model atmospheres computed 
by the code STERNE (Jeffery, Woolf \& Pollacco 2001).
STERNE model atmosphere was combined with the Armagh LTE code 
SPECTRUM (Jeffery, Woolf \& Pollacco 2001) to compute the equivalent width of a line
or a synthetic spectrum. In matching
a synthetic spectrum to an observed spectrum we include broadening
due to the instrumental profile, the microturbulent velocity $\xi$ and assign all
additional broadening, if any, to rotational broadening.
We use the standard rotational broadening
function $V(v\sin i,\beta)$ with
the limb darkening coefficient set at $\beta = 1.5$ (Jeffery, Woolf \& Pollacco 2001).
Observed unblended line profiles are used to obtain the projected rotational
velocity $v\sin i$. Synthetic line profile, including
the broadening due to instrumental profile, for
the adopted model atmosphere ($T_{\rm eff}$,$\log g, \xi$) and the abundance,
is found to be sharper than the observed. This extra broadening in the observed profile
is attributed to rotational broadening. 

The adopted $gf$-values for C, N, O, are from Wiese, Fuhr \& Deters (1996), and
for rest of the elements are from 
NIST database\footnote{http://physics.nist.gov/PhysRefData/ASD/lines\_form.html},
Kurucz's database\footnote{http://kurucz.harvard.edu}, Th\'{e}venin (1989, 1990),
and the compilations by R. E. Luck (private communication).
The Stark broadening and radiative broadening coefficients, if available, are mostly
taken from the Kurucz's database. 
The data  for
computing He\,{\sc i} profiles are the same as in Jeffery, Woolf \& Pollacco (2001).
Table A1 has the detailed line list used in our analysis.

\subsection{Atmospheric parameters}

The model atmospheres are defined by the effective temperature, the
surface gravity, and the chemical composition. The input composition of
He and C (the C/He ratio) is fully consistent with the C/He ratio derived from
the observed spectrum with that model; He and C abundances, particularly He,
dominate the continuous opacity. The input composition of rest of the elements
is solar with H/He fixed at 10$^{-4}$ by number.

The analysis involves the determination of effective temperature ($T_{\rm eff}$),
surface gravity ($\log g$), and microturbulent velocity ($\xi$) before estimating
the photospheric elemental abundances of the star. These parameters are determined 
from the line spectrum. The microturbulent velocity $\xi$ (in km s$^{-1}$) is first 
determined by the requirement that the abundance from
a set of lines of the same ion with similar excitation potentials be independent of 
a line's equivalent width.
 For an element represented in the
spectrum by two or more ions, imposition of ionization
equilibrium (i.e., the same abundance is required from lines of
different stages of ionization) defines a locus in the
($T_{\rm eff},\log g)$ plane.
 Different pairs of ions of a common element provide
loci of very similar slope in the ($T_{\rm eff},\log g)$ plane.

An indicator yielding a locus with a contrasting slope
in the ($T_{\rm eff},\log g)$ plane is required to break the
degeneracy presented by ionization equilibria.
A potential indicator is a He\,{\sc i} line.
For stars hotter than about 10,000 K,
the  He\,{\sc i} lines are less sensitive to $T_{\rm eff}$
than to $\log g$ on account of pressure broadening due to
the quadratic Stark effect.
The diffuse series lines are, in particular, useful because they are
less sensitive to the microturbulent velocity than the sharp lines.
A second indicator may be available: species represented
by lines spanning a range in excitation potential may serve as
a thermometer measuring $T_{\rm eff}$ with a weak dependence
on $\log g$.

\section{LSS\,3378 - abundance analysis results}
 
First to determine is the microturbulent velocity $\xi$. We adopt a model atmosphere
with $T_{\rm eff}$ = 10000 K, which is close to that found by Jeffery et al. (2001),
and adopt $\log g$ = 1.0
which is a fair assumption for these hydrogen-deficient supergiants.  $\xi$ is found
to be 4.5 and 7.5 km s$^{-1}$ from Fe\,{\sc ii} and C\,{\sc i} lines, respectively.
We adopt $\xi$ = 6 km s$^{-1}$ for abundance determination, and Figure 1 illustrates 
the method for obtaining $\xi$. Fe\,{\sc ii} lines used for determining $\xi$ were of 
similar lower excitation potential and so was the case for C\,{\sc i} lines. 

\begin{figure}
\epsfxsize=8truecm
\epsffile{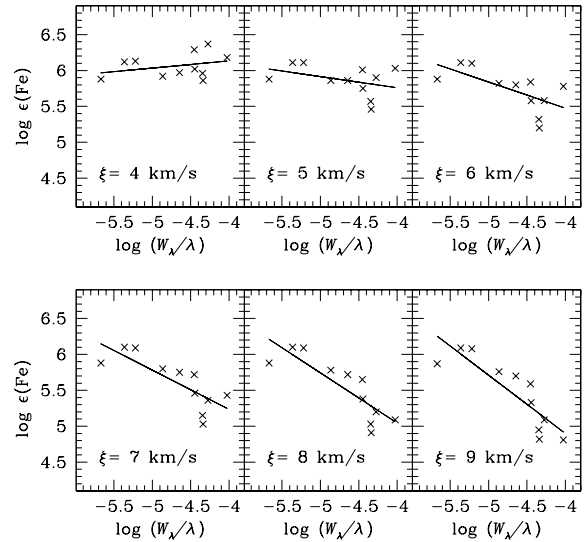}
\caption{Abundances from Fe\,{\sc ii} lines for LSS\,3378 versus their
reduced equivalent widths (log $W_{\lambda}/\lambda$).
A microturbulent velocity of $\xi \simeq 4.5$ km s$^{-1}$ is obtained from this
figure.}
\end{figure}

\begin{figure}
\epsfxsize=8truecm
\epsffile{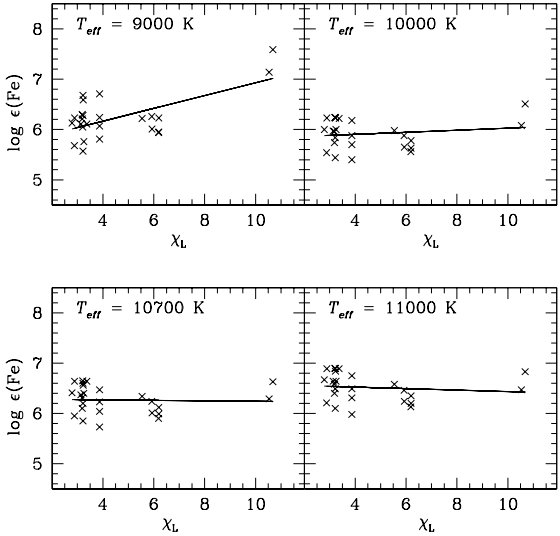}
\caption{Excitation equlilbrium for LSS\,3378 using Fe\,{\sc ii} lines, and
models with $\log g$ = 0.5.}
\end{figure}

\begin{figure}
\epsfxsize=8truecm
\epsffile{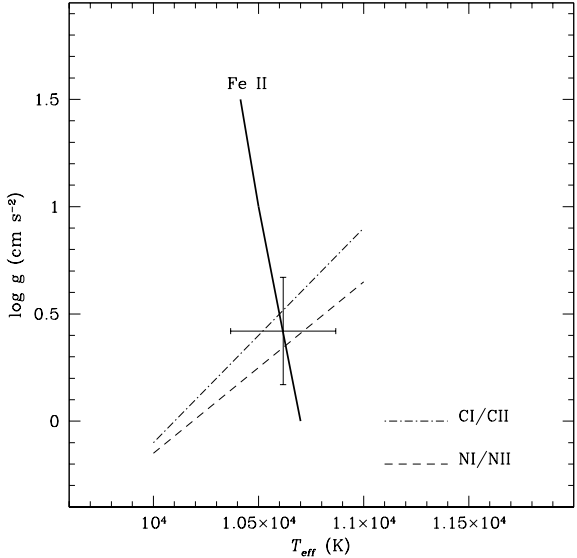}
\caption{The $T_{\rm eff}$ vs $\log g$ plane for LSS\,3378. Loci satisfying
ionization equilibria are plotted -- see key on the figure.
The locus satisfying the excitation balance of Fe\,{\sc ii}
lines is shown by thick solid line.
The cross shows the adopted model atmosphere parameters.}
\end{figure}
 
\begin{figure}
\epsfxsize=8truecm
\epsffile{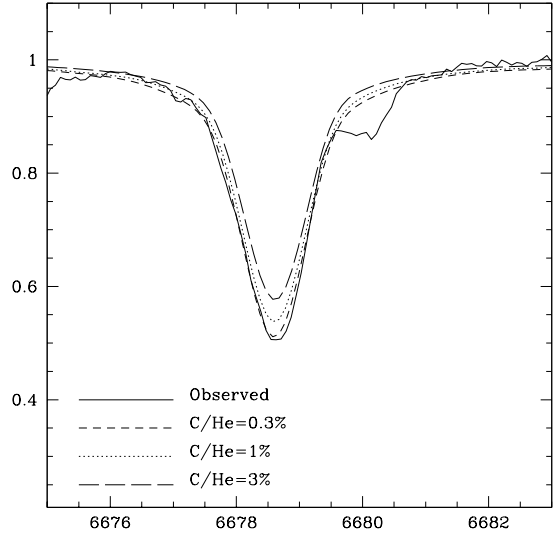}
\caption{LSS\,3378's observed and synthesized He\,{\sc i} line profile
at 6678.15\AA. The He\,{\sc i} line profiles are synthesized using the model
$T_{\rm eff}$ = 10600 K and $\log g$ = 0.4, for different values of C/He - see key
on the figure.} 
\end{figure}

 The $T_{\rm eff}$ was estimated from Fe\,{\sc ii} lines spanning excitation 
potentials from 3 eV to 11 eV. For $\xi$ = 4.5 km s$^{-1}$,
models ($T_{\rm eff}$, $\log g$) were found which gave the same abundance independent
of excitation potential. The $T_{\rm eff}$ determined from Fe\,{\sc ii} lines
somewhat depends on the adopted surface gravity. 
Figure 2 illustrates the method for obtaining $T_{\rm eff}$.
Ionization equilibrium loci 
for C\,{\sc i}/C\,{\sc ii} and N\,{\sc i}/N\,{\sc ii} are shown in Figure 3.
These with the $T_{\rm eff}$ 
determined from the excitation equilibrium of Fe\,{\sc ii} lines, which is a function 
of surface gravity, gives the estimate of the
stellar parameters: $T_{\rm eff}$ = 10600$\pm$250 K, $\log g$ = 0.4$\pm$0.25 cgs, and
$\xi$ = 6$\pm$2 km s$^{-1}$ (Figure 3). Thus, the abundance analysis was conducted
for the model atmosphere (10600,0.4,6.0).
At this $T_{\rm eff}$ which is close to 11000 K, electron scattering (most of
the free electrons from photoionization of neutral helium), and photoionization of
neutral helium are the major sources of continuous opacity in the line forming
regions (Pandey et al. 2001).  Hence, for LSS\,3378 it is evident that helium controls
the continuum opacity, and the C/He ratio is directly determined 
from the measured equivalent widths of C\,{\sc i} and C\,{\sc ii}
lines observed in the spectrum of LSS\,3378.
The C\,{\sc i} lines give log $\epsilon$(C) = 9.53$\pm$0.19. Four C\,{\sc ii}
lines give 9.44$\pm$0.11. The abundances from C\,{\sc i} and C\,{\sc ii} lines
are in good agreement, and imply a C/He = 1\%.
At the $T_{\rm eff}$ of LSS\,3378, the He\,{\sc i} lines are mildly sensitive to the
C/He ratio. The C/He ratio may be derived by
fitting the He\,{\sc i} lines at 5048, 5876, and 6678\AA\ (Figure 4); the blending
Ne\,{\sc i} line at 6678\AA\ is taken into account.
These He\,{\sc i} lines give C/He = 0.8$\pm$0.1\%, where the uncertainty reflects only the scatter
of the three results.
LSS\,3378's final abundances, for key elements, derived for C/He = 1\% are 
summarized in Table 1; also given are solar abundances from Table 2 of Lodders (2003) 
for comparison.
The individual elemental abundances listed in Table 1 are given as
log $\epsilon(i)$, normalized such that log $\Sigma$$\mu_i \epsilon(i)$ = 12.15
where $\mu_i$ is the atomic weight of element $i$.
The lines used for abundance analysis including the mean abundance, and the
line-to-line scatter are given in Table A1. The abundance errors due to the
uncertainty in the adopted stellar parameters are given in Table 2.
The deduced $v\sin i$ is about 26 km s$^{-1}$.

\begin{table*}
\caption{Adopted key elemental abundances for LSS\,3378}
\begin{center}
%{\small
\begin{tabular}{lrcccccccccccccccc} \hline
Star & H & He & C & N & O & Mg & Al & Si & P & S & Ti & Cr & Mn & Fe & Y & Zr & Ba \\
\hline
LSS\,3378 & 7.18 & 11.50 & 9.46 & 8.26 & 9.29 & 5.97 & 5.98 & 6.62 & 4.85 & 6.52 & 4.29 & 4.47 &
4.57 & 6.11 & 2.85 & 3.51 & 2.49 \\
Sun       & 12.00& 10.98 & 8.46 & 7.90 & 8.76 & 7.62 & 6.54 & 7.61 & 5.54 & 7.26 & 5.00 & 5.72 & 
5.58 & 7.54 & 2.28 & 2.67 & 2.25 \\
\hline
\end{tabular}%}
\end{center}
\end{table*}

\section{Abundances}

%The derived abundances of LSS\,3378 are very similar to other EHes and
%this star belongs to the O-rich group of EHes (see Figures ...).
The derived abundances of the EHe LSS\,3378 are compared with the measured
abundances, taken from the literature, of the other EHes (see Pandey et al. 2006).
The photospheric composition of LSS\,3378 reveals that the surface material
is contaminated by the products of H-burning, and He-burning reactions, as observed
for most of the EHes.

LSS\,3378's abundance ratios: Cr/Fe, Mn/Fe, S/Fe, Si/Fe, and Ti/Fe, are
as expected for that in metal-poor normal and unevolved stars except for the low
Mg/Fe ratio. The abundances
of Fe, Cr, and Mn$-$the iron peak elements, and S, Si, and Ti$-$the 
$\alpha$-elements represent the initial metallicity of LSS\,3378 as these elements are 
expected to be unaffected by H- and the He-burning, and
attendant nuclear reactions.
We choose Fe to be the indicator of the initial metallicty in
LSS\,3378 for spectroscopic convenience.

LSS\,3378's derived elemental abundances that are affected by evolution
are of H, C, N, O, Ne, Y, Zr, and Ba. Note that, the Ne abunadnaces from 
Ne\,{\sc i} lines are affected by
departures from LTE, hence, not discussed here. However, the neon LTE abunadnces
from Ne\,{\sc i} lines are listed in Table A1. \\
{\it Hydrogen}$-$H abundance log $\epsilon$(H) is about 7.2 that fits the suggested trend
of increasing H with increasing $T_{\rm eff}$ for all the EHes with C/He ratio of
about 1\%, the exception being the hottest EHe 
LS IV $+6^\circ2$ (see Pandey et al. 2006).\\
{\it Carbon}$-$The C/He ratio is 0.0083, the mean C/He ratio from 15 EHes, that excludes
the two EHes HD\,144941 and V652\,Her with much lower C/He ratio, is 0.0066.\\
{\it Nitrogen}$-$Nitrogen is enriched above its initial abundance expected based
on the Fe abundance. The observed N abundance is the result of complete
conversion of the initial C, N, and O to N via H-burning CN-cycle and the ON-cycles.\\
{\it Oxygen}$-$Oxygen abundance relative to Fe is overabundant by about 1.5 dex.
LSS\,3378 fits to the oxygen-rich group of EHes (see Pandey et al. 2006).\\
{\it Yttrium, Zr, {\rm and} Ba}$-$Relative to iron, Y, Zr, and Ba are overabundant with
respect to solar by about a factor of 80 (1.9 dex), 150 (2.2 dex), and 40 (1.6 dex),
respectively. Note that, the measured abundances are are based on one or two lines.
The observed Y, Zr, and Ba overabundances are attributed to contamination of the
atmosphere by $s$-process products.

\section{Conclusions}

The abundances of the cool EHe star LSS\,3378 are measured. The measured C/He ratio
is about 1\% similar to most of the EHes. LSS\,3378 fits the oxygen-rich group
of EHes discussed by Pandey et al. (2006). The measured abundances of Y, Zr, and Ba
suggest that in the EHe star LSS\,3378, $s$-process nucleosynthesis did occur in its
earlier evolution. With this analysis, a total of four EHes: LSS\,3378, PV\,Tel,
V1920\,Cyg, and LSE\,78, show a strong enhancement of Y and Zr attributable to an
$s$-process (Pandey et al. 2004, 2006).

An interesting similarity is suggested by the $s$-process abundances in LSS\,3378
and those of the RCB stars. The light $s$-process elements are more enhanced than the
heavy $s$-process elements in RCB stars (Asplund et al. 2000; Rao \& Lambert 2003;
Vanture, Zucker, \& Wallerstein 1999; Bond, Luck, \& Newman 1979). 
Similar enhancement of light $s$-process elements (Y, Zr) over heavy $s$-process 
element (Ba) is observed in LSS\,3378.

Enrichment of $s$-process elements is not expected for the EHes resulting from
a merger of a He with a C-O white dwarf as discussed by Pandey et al. (2004, 2006).
However, synthesis by neutrons via the $s$-process may occur during the merger and
needs to be explored.
 
To further improve the chemical analysis, non-LTE calculations should be performed
for key elements like N, O, Ne, Si, S, and Fe, and in particular neon.

\begin{table}
%\centering
%\begin{minipage}{75mm}
\caption{Abundance errors due to uncertainties in the stellar parameters: $\Delta$$T_{\rm eff}$,
$\Delta$$\log g$, and $\Delta$$\xi$. The abundance error due to $\Delta$$T_{\rm eff}$ is the difference
in abundances derived from the adopted model ($T_{\rm eff}$, $\log g$, $\xi$) and 
a model ($T_{\rm eff}$+$\Delta$$T_{\rm eff}$, $\log g$, $\xi$). The abundance error due to
$\Delta$$\log g$ is the difference in abundances derived from the adopted
model ($T_{\rm eff}$, $\log g$, $\xi$) and a model ($T_{\rm eff}$, $\log g$+$\Delta$$\log g$, $\xi$).
The abundance error due to $\Delta$$\xi$ is the difference in abundances derived from the adopted
model ($T_{\rm eff}$, $\log g$, $\xi$) and a model ($T_{\rm eff}$, $\log g$, $\xi$+$\Delta$$\xi$).}
\begin{tabular}{lccc} \hline
\multicolumn{1}{l}{Species}&\multicolumn{1}{c}{$\Delta$$T_{\rm eff}$ = +300}&
\multicolumn{1}{c}{$\Delta$$\log g$ = +0.25}& 
\multicolumn{1}{c}{$\Delta$$\xi$ = +2.0} \\
\multicolumn{1}{c}{}& \multicolumn{1}{c}{[K]} &
\multicolumn{1}{c}{[cgs]} &\multicolumn{1}{c}{[km s$^{-1}$]}\\
\hline
H\,{\sc i} & $-$0.19 & +0.22 & +0.09\\
C\,{\sc i} & $-$0.18 & +0.18 & +0.11\\
C\,{\sc ii} & +0.02 & +0.02 & +0.13\\
N\,{\sc i} & $-$0.18 & +0.18 & +0.12\\
N\,{\sc ii} & +0.04 & $-$0.02 & +0.09\\
O\,{\sc i} & $-$0.17 & +0.20 & +0.24\\
Ne\,{\sc i} & $-$0.04 & +0.11 & +0.41\\
Na\,{\sc i} & $-$0.17 & +0.17 & +0.09\\
Mg\,{\sc ii} & $-$0.22 & +0.18 & +0.28\\
Al\,{\sc ii} & $-$0.12 & +0.11 & +0.17\\
Al\,{\sc iii} & +0.04 & $-$0.05 & +0.09\\
Si\,{\sc ii} & $-$0.15 & +0.16 & +0.26\\
P\,{\sc ii} & $-$0.07 & +0.03 & +0.05\\
S\,{\sc ii} & 0.00 & 0.00 & +0.20\\
%Sc\,{\sc ii} & $-$0.26 & +0.18 & +0.01\\
Ti\,{\sc ii} & $-$0.27 & +0.18 & +0.03\\
Cr\,{\sc ii} & $-$0.35 & +0.17 & $-$0.05\\
Mn\,{\sc ii} & $-$0.22 & +0.17 & +0.03\\
Fe\,{\sc ii} & $-$0.25 & +0.15 & +0.11\\
Y\,{\sc ii} & $-$0.27 & +0.19 & +0.01\\
Zr\,{\sc ii} & $-$0.28 & +0.19 & $-$0.01\\
Ba\,{\sc ii} & $-$0.25 & +0.19 & +0.01\\
\hline
\end{tabular}
%\end{minipage}
\end{table}

The observations presented here were obtained at CTIO, National
Optical Astronomy Observatories (NOAO), which is operated by the Association of
Universities for Research in Astronomy Inc. (AURA) under a cooperative agreement
with the National Science Foundation, USA. We thank the referee Tony Lynas-Gray
for drawing our attention to the strong Na\,D profiles in LSS\,3378's spectrum.

\appendix

\section{Lines used for abundance analysis}

The lines used for the abundance analysis of LSS\,3378
are given in Table A1. The lower excitation potential ($\chi_L$),
$gf$-value, measured equivalent width ({\em $W_{\lambda}$}) and the 
abundance (log $\epsilon$) derived for each line is also
listed. The abundances are derived using $C/He$ = 1.0\% model for 
LSS\,3378.

\begin{table}
\caption{Lines used to derive elemental abundances for LSS\,3378}
\begin{tabular}{lrrrrl}
\hline
Ion&&&{\em W$_{\lambda}$}&&\\
$\lambda$ (\AA)& $\chi$ (eV)& log $gf$& (m\AA)& log $\epsilon$$^a$
& Ref.$^b$\\
\hline
H\,{\sc i}&&&&&\\
%No. of lines=1&&&&&\\
   6562.8&    10.20&   0.71&    599&    7.18& Jeffery \\
\hline
He\,{\sc i}&&&&&\\
%No. of lines=3&&&&&\\
% 3888.65&   19.73&   -0.70&    368&   11.59& Jeffery\\
% 3926.53&   21.13&  -1.65&    185&   11.75& Jeffery\\
% 3964.73&   20.61&   -1.30&    210&   11.39& Jeffery\\
% 4120.81&   20.87&  -1.52&    200&   11.58& Jeffery\\
% 4471.48&   20.87&   0.04&    375&   11.33& Jeffery\\
% 4713.14&   20.87&  -1.07&    267&   11.73& Jeffery\\
  5047.74&   21.22&  $-$1.60&   Synth$^c$&11.54   &Jeffery \\
  5875.62&   20.96&   0.74&   Synth&11.54   & Jeffery \\
  6678.15&   21.22&   0.33&   Synth&11.54   & Jeffery \\
\hline
C\,{\sc i}&&&&&\\
%No. of lines=30&&&&&\\
5052.18 & 7.69 & $-$1.30 &252 &   9.26 &  Wiese \\
5380.34 & 7.69 & $-$1.62 &193 &   9.13 &  Wiese \\
5540.76 & 8.64 & $-$2.40 & 81 &   9.59 &  Wiese \\
5817.70 & 8.85 & $-$2.86 & 19 &   9.41 &  Wiese \\
%5850.24 & 8.77 & $-$2.70 & 36 &   9.51 &  Wiese \\
%5963.99 & 8.65 & $-$2.70 & 81 &   9.90 &  Wiese \\
6001.12 & 8.64 & $-$2.05 &164 &   9.83 &  Wiese \\
6002.98 & 8.65 & $-$2.16 & 83 &   9.37 &  Wiese \\
6010.68 & 8.64 & $-$1.96 &166 &   9.75 &  Wiese \\
6014.83 & 8.64 & $-$1.59 &215 &   9.72 &  Wiese \\
6078.39 & 8.85 & $-$2.27 & 74 &   9.51 &  Wiese \\
6113.15 & 8.85 & $-$2.63 & 66 &   9.79 &  Wiese \\
6115.84 & 8.85 & $-$2.52 & 85 &   9.85 &  Wiese \\
6120.81 & 8.85 & $-$2.40 & 84 &   9.72 &  Wiese \\
%6337.18 & 8.77 & $-$2.30 & 59 &   9.36 &  Wiese \\
%6342.32 & 8.77 & $-$2.10 & 74 &   9.29 &  Wiese \\
6413.55 & 8.77 & $-$2.00 &153 &   9.78 &  Wiese \\
6568.71 & 9.00 & $-$2.16 & 93 &   9.62 &  Wiese \\
6587.61 & 8.54 & $-$1.00 &268 &   9.43 &  Wiese \\
6591.46 & 8.85 & $-$2.40 & 54 &   9.47 &  Wiese \\
6595.24 & 8.85 & $-$2.40 & 35 &   9.24 &  Wiese \\
6611.35 & 8.85 & $-$1.85 &123 &   9.47 &  Wiese \\
6671.85 & 8.85 & $-$1.66 &166 &   9.55 &  Wiese \\
6688.79 & 8.85 & $-$2.16 & 95 &   9.56 &  Wiese \\
6711.32 & 8.54 & $-$2.70 & 50 &   9.57 &  Wiese \\
7108.93 & 8.64 & $-$1.59 &182 &   9.48 &  Wiese \\
7111.47 & 8.64 & $-$1.09 &252 &   9.46 &  Wiese \\
7113.18 & 8.65 & $-$0.77 &288 &   9.38 &  Wiese \\
7476.18 & 8.77 & $-$1.57 &193 &   9.60 &  Wiese \\
7685.19 & 8.77 & $-$1.52 &182 &   9.48 &  Wiese \\
7832.64 & 8.85 & $-$1.82 &123 &   9.43 &  Wiese \\
7860.88 & 8.85 & $-$1.15 &241 &   9.48 &  Wiese \\
%\end{tabular}
%}
%\end{center}
%\newpage
%\begin{center}
%{\bf Table 4.6} (continued)\\
%\vspace{0.5cm}
%{\it Lines used to derive elemental abundances for FQ Aqr}\\
%\vspace{0.5cm}
%\noindent
%{\scriptsize
%\begin{tabular}{lccccl}
%\hline \hline
%$\lambda$(\AA~)& $\chi$(eV)& log gf& W(m\AA~)& Abundance$^{a}$& Reference$^{b}$\\
%\hline
\hline
Mean:&&&&9.53$\pm$0.18&\\
\hline
C\,{\sc ii}&&&&&\\
%No. of lines=2&&&&&\\
5537.61 & 19.50 & $-$1.79 & 42 & 9.57 &  Wiese \\
5891.60 & 18.05 & $-$0.44 &204 & 9.34 &  Wiese\\
6578.05 & 14.45 & $-$0.03 &848 & 9.36 &  Wiese\\
6582.88 & 14.45 & $-$0.33 &728 & 9.48 &  Wiese\\
\hline
Mean:&&&& 9.44$\pm$0.11&\\
\hline
\end{tabular}
%\medskip
%$^a$Normalized such that log$\Sigma \mu_{i} \epsilon(i)$ = 12.15\\
%$^b$Sources of $gf$-values
\end{table}
\begin{table}
\contcaption{}
\begin{tabular}{lrrrrl}
\hline
Ion&&&{\em W$_{\lambda}$}&&\\
$\lambda$ (\AA)& $\chi$ (eV)& log $gf$&  (m\AA)& log $\epsilon$$^a$
& Ref.$^b$\\
\hline
N\,{\sc i}&&&&&\\
%No. of lines=5&&&&&\\
6008.46 & 11.60 & $-$1.11 & 75  & 8.30  & Wiese \\
6644.96 & 11.76 & $-$0.86 & 63  & 8.03  & Wiese \\
7406.22 & 12.01 & $-$0.74 & 68  & 8.08  & Wiese \\
7423.64 & 10.33 & $-$0.71 &237  & 8.32  & Wiese \\
7442.30 & 10.33 & $-$0.38 &302  & 8.40  & Wiese \\
\hline
Mean:&&&&8.23$\pm$0.16&\\
\hline
N\,{\sc ii}&&&&&\\
%No. of lines=2&&&&&\\
5045.10 & 18.48 & $-$0.41 &  79  &  8.37 & Wiese  \\  
5666.63 & 18.47 & $-$0.05 &  86  &  8.14 & Wiese  \\
5679.56 & 18.48 &  0.25 & 118  &  8.17 & Wiese  \\
5686.21 & 18.47 & $-$0.55 &  38  &  8.05 & Wiese  \\
5710.77 & 18.48 & $-$0.52 &  79  &  8.55 & Wiese  \\
\hline
Mean:&&&&8.26$\pm$0.20&\\
\hline
O\,{\sc i}&&&&&\\
%No. of lines=8&&&&&\\
4968.79 & 10.74 & $-$1.26  &  233   &  9.35 &   Wiese \\
5330.74 & 10.74 & $-$0.88  &  302   &  9.35 &   Wiese \\
5436.86 & 10.74 & $-$1.40  &  208   &  9.25 &   Wiese \\
5555.00 & 10.99 & $-$1.80  &  149   &  9.35 &   Wiese \\
6158.19 & 10.74 & $-$0.30  &  401   &  9.41 &   Wiese \\
7473.24 & 14.12 & $-$0.37  &   86   &  9.00 &   Wiese \\
\hline
Mean:&&&&9.29$\pm$0.15&\\
\hline
Ne\,{\sc i}&&&&&\\
%No. of lines=14&&&&&\\
%5748.30 & 18.56 & $-$1.02  &  119   &  8.90  &  NIST \\
%5764.42 & 18.56 & $-$0.31  &  200   &  8.98  &  NIST \\
5852.49 & 16.85 & $-$0.46  &  302   &  9.21  &  NIST \\
5881.90 & 16.62 & $-$0.75  &  280   &  9.20  &  NIST \\
%5944.83 & 16.62 & $-$0.52  &  319   &  9.30  &  NIST \\
%5965.47 & 18.73 & $-$1.13  &  133   &  9.25  &  NIST \\
6030.00 & 16.67 & $-$1.04  &  231   &  9.08  &  NIST \\
6074.34 & 16.67 & $-$0.48  &  338   &  9.43  &  NIST \\
6143.06 & 16.62 & 0.10   &  505   &  9.93  &  NIST \\
6163.59 & 16.72 & $-$0.60  &  315   &  9.39  &  NIST \\
6217.28 & 16.62 & $-$0.96  &  247   &  9.11  &  NIST \\
6266.50 & 16.72 & $-$0.36  &  404   &  9.78  &  NIST \\
6334.43 & 16.62 & $-$0.32  &  419   &  9.76  &  NIST \\
6382.99 & 16.67 & $-$0.23  &  446   &  9.85  &  NIST \\
6402.25 & 16.62 & 0.35   &  584   &  9.74  &  NIST \\
6506.53 & 16.67 &$-$0.02   &  485   &  9.80  &  NIST \\
6598.95 & 16.85 & $-$0.34  &  357   &  9.55  &  NIST \\
7032.41 & 16.62 & $-$0.25  &  449   &  9.84  &  NIST \\
%\end{tabular}
%}
%\end{center}
% 
%\newpage
% 
%\begin{center}
%{\bf Table} (continued)\\
%\vspace{0.5cm}
%{\it Lines used to derive elemental abundances for FQ Aqr}\\
%\vspace{0.5cm}
%\noindent
%{\scriptsize
%\begin{tabular}{lccccl}
%\hline
%$\lambda$(\AA~)& $\chi$(eV)& log gf& W(m\AA~)& Abundance$^{a}$& Reference$^{b}$\\
%\hline
\hline
Mean:&&&&9.55$\pm$0.30&\\
\hline
Na\,{\sc i}&&&&&\\
%No. of lines=2&&&&&\\
5682.63 & 2.10 & $-$0.70  &  32  &  6.40  &  NIST \\
6160.75 & 2.10 & $-$1.23  &  29  &  6.89  &  NIST \\
8194.82 & 2.10 &  0.51  & 257  &  6.78  &  NIST \\
\hline
Mean:&&&&6.69$\pm$0.26&\\
\hline
Mg\,{\sc ii}&&&&&\\
%No. of lines=6&&&&&\\
7877.05 & 10.00 & 0.39  & 246  & 5.89  &  NIST \\  
7896.37 & 10.00 & 0.65  & 310  & 6.04  &  NIST \\
\hline
Mean:&&&&5.97$\pm$0.11&\\
\hline
\end{tabular}
%\medskip
%$^a$Normalized such that log$\Sigma \mu_{i} \epsilon(i)$ = 12.15\\
%$^b$Sources of $gf$-values
\end{table}
\begin{table}
\contcaption{}
\begin{tabular}{lrrrrl}
\hline
Ion&&&{\em W$_{\lambda}$}&&\\
$\lambda$ (\AA)& $\chi$ (eV)& log $gf$& (m\AA)& log $\epsilon$$^a$
& Ref.$^b$\\
\hline
Al\,{\sc ii}&&&&&\\
%No. of lines=4&&&&&\\
5593.23 & 13.26 & 0.41 &   95  &  5.19 & NIST  \\ 
6226.13 & 13.07 & 0.05 &   86  &  5.39 & NIST  \\
6231.72 & 13.07 & 0.40 &  171  &  5.71 & NIST  \\
6823.40 & 13.07 &$-$0.14 &  102  &  5.72 & NIST  \\
6837.08 & 13.08 & 0.08 &  122  &  5.66 & NIST  \\
7042.05 & 11.32 & 0.35 &  393  &  6.52 & NIST  \\
%7063.64 & 11.32 &$-$0.35 &  289  &  6.43:& NIST  \\
\hline
Mean:&&&&5.70$\pm$0.45&\\
\hline
Al\,{\sc iii}&&&&&\\
5696.47 & 15.64&  0.24  &  85  &  6.04 &  NIST  \\  
5722.65 & 15.64& $-$0.07  &  57  &  6.03 &  NIST  \\
\hline
Mean:&&&&6.04$\pm$0.01&\\
\hline
Si\,{\sc ii}&&&&&\\
%No. of lines=6&&&&&\\
5055.98 & 10.07 &  0.44 & 490  &  6.86 &  NIST \\   
5957.56 & 10.07 & $-$0.35 & 304  &  6.72 &  NIST \\
5978.93 & 10.07 & $-$0.06 & 329  &  6.59 &  NIST \\
6829.83 & 12.88 & $-$0.27 &  69  &  6.31 &  NIST \\
\hline
Mean:&&&&6.62$\pm$0.23&\\
\hline
P\,{\sc ii}&&&&&\\
%No. of lines=2&&&&&\\
5499.69 & 10.80 & $-$0.30  &  38  & 4.69 & NIST \\  
6024.13 & 10.76 & 0.14   & 111  & 4.94 & NIST \\
6034.04 & 10.74 & $-$0.22  &  71  & 4.95 & NIST \\
6165.57 & 10.80 & $-$0.34  &  45  & 4.83 & NIST \\
\hline
Mean:&&&&4.85$\pm$0.12&\\
\hline
S\,{\sc ii}&&&&&\\
%No. of lines=3&&&&&\\
5009.52  &13.62 & $-$0.28  & 169 & 6.81  & NIST \\   
5014.00  &14.07 & 0.10   & 173 & 6.70  & NIST \\
5103.29  &13.67 & $-$0.11  & 108 & 6.11  & NIST \\
5320.70  &15.07 & 0.49   & 151 & 6.56  & NIST \\
5428.64  &13.58 & $-$0.13  & 176 & 6.72  & NIST \\
%5432.74 & 13.62&  0.26  &  292&  7.31 &  NIST \\
%5453.79 & 13.67&  0.48  &  326&  7.34 &  NIST \\
5473.60  &13.58 & $-$0.18  & 154 & 6.57  & NIST \\
%5509.65 & 13.62&  $-$0.14 &  207&  7.03 &  NIST \\
5555.97  &13.62 & $-$0.99  &  45 & 6.32  & NIST \\
5606.09  &13.73 & 0.31   & 194 & 6.54  & NIST \\
5616.61  &13.66 & $-$0.64  &  88 & 6.46  & NIST \\
5659.96  &13.68 & $-$0.05  & 170 & 6.62  & NIST \\
5664.76  &13.66 & $-$0.25  & 111 & 6.29  & NIST \\
\hline
Mean:&&&&6.52$\pm$0.21&\\
\hline
%Sc\,{\sc ii}&&&&&\\
%No. of lines=7&&&&&\\
%5684.20 & 1.51 & $-$1.25 &23.0 & 3.58 &  Luck \\
%\hline
Ti\,{\sc ii}&&&&&\\
5072.28 & 3.12 & $-$0.75 &   36.0  &  4.18 & NIST \\  
5188.68 & 1.58 & $-$1.21 &  102.0  &  4.39 & NIST \\
\hline
Mean:&&&&4.29$\pm$0.15&\\
\hline
Cr\,{\sc ii}&&&&&\\
5246.77 & 3.71 & $-$2.48 &  18.0 &  4.42 &  NIST \\   
%5274.96 & 4.07 & $-$1.29 & 170.0 &  4.76 &  Kurucz \\
5313.56 & 4.07 & $-$1.65 &  90.0 &  4.62 &  NIST \\
%5334.87 & 4.07 & $-$1.56 & 135.0 &  4.82 &  Kurucz \\
5420.92 & 3.76 & $-$2.58 &  20.0 &  4.37 &  NIST \\
%5478.37 & 4.18 & $-$1.91 &  68.0 &  4.76 &  Kurucz \\
\hline
Mean:&&&&4.47$\pm$0.13&\\
\hline
\end{tabular}
%\medskip
%$^a$Normalized such that log$\Sigma \mu_{i} \epsilon(i)$ = 12.15\\
%$^b$Sources of $gf$-values
\end{table}
\begin{table}
\contcaption{}
\begin{tabular}{lrrrrl}
\hline
Ion&&&{\em W$_{\lambda}$}&&\\
$\lambda$ (\AA)& $\chi$ (eV)& log $gf$& (m\AA)& log $\epsilon$$^a$
& Ref.$^b$\\
\hline
Mn\,{\sc ii}&&&&&\\
%No. of lines=3&&&&&\\
5297.06&  9.86 & 0.87  &  47.0 &  4.67  & Kurucz \\   
6122.45& 10.18 & 0.95  &  57.0 &  4.85  & Kurucz \\
7415.81&  3.71 &$-$2.20  &  40.0 &  4.37  & Kurucz \\
7432.30&  3.71& $-$2.50  &  21.0 &  4.37  & Kurucz \\
\hline
Mean:&&&&4.57$\pm$0.24&\\
\hline
Fe\,{\sc ii}&&&&&\\
%No. of lines=59&&&&&\\
4993.36 & 2.81 & $-$3.65   &  113.0  &  6.32  & NIST  \\  
%5081.90 & 10.38&  $-$0.59  &   87.0  &  7.04  & NIST  \\
5169.03 & 2.89 & $-$0.87   &  487.0  &  6.28  & NIST  \\
5197.58 & 3.23 & $-$2.10   &  235.0  &  5.77  & NIST  \\
5234.63 & 3.22 & $-$2.05   &  278.0  &  6.02  & NIST  \\
5247.95 & 10.53&  0.63   &  121.0  &  6.14  & NIST  \\
5254.93 & 3.23 & $-$3.00   &   97.0  &  5.77  & Kurucz \\
5272.40 & 5.96 & $-$2.05   &   82.0  &  6.18  & NIST  \\
5276.00 & 3.20 & $-$1.96   &  242.0  &  5.64  & NIST  \\
5284.11 & 2.89 & $-$3.19   &  110.0  &  5.86  & NIST  \\
%5325.55 & 3.22 & $-$2.52   &  105.0  &  5.34  & NIST  \\
5362.87 & 3.20 & $-$2.70   &  192.0  &  6.03  & Kurucz  \\
5525.13 & 3.27 & $-$4.61   &   24.0  &  6.64  & Kurucz  \\
5534.85 & 3.25 & $-$2.93   &  195.0  &  6.29  & NIST  \\
5591.37 & 3.27 & $-$4.69   &   12.0  &  6.40  & Kurucz  \\
5627.50 & 3.39 & $-$4.36   &   33.0  &  6.62  & NIST  \\
5952.51 & 5.96 & $-$2.05   &   61.0  &  5.97  & Kurucz  \\
5961.71 & 10.68&  0.69   &  162.0  &  6.41  & Kurucz  \\
5991.38 & 3.15 & $-$3.74   &   82.0  &  6.32  & NIST  \\
6147.74 & 3.89 & $-$2.70   &  123.0  &  5.94  & Kurucz  \\
6149.26 & 3.89 & $-$2.92   &  118.0  &  6.13  & NIST  \\
6175.15 & 6.22 & $-$2.00   &   64.0  &  6.08  & Kurucz  \\
6179.38 & 5.57 & $-$2.81   &   42.0  &  6.32  & NIST  \\
6247.56 & 3.89 & $-$2.52   &  199.0  &  6.21  & NIST  \\
6331.96 & 6.22 & $-$1.96   &   48.0  &  5.87  & Kurucz  \\
6446.41 & 6.22 & $-$2.16   &   38.0  &  5.97  & NIST  \\
7462.41 & 3.89 & $-$2.70   &   90.0  &  5.68  & Kurucz \\
\hline
Mean:&&&&6.11$\pm$0.27&\\
\hline
Y\,{\sc ii}&&&&&\\
%No. of lines=3&&&&&\\
5205.73 & 1.03 & $-$0.34  &   28.0  &  2.85  & Luck \\
\hline
Zr\,{\sc ii}&&&&&\\
4962.29 & 0.97 & $-$1.69  &    6.0  &  3.42 & Thevenin \\  
5350.09 & 1.83 & $-$0.93  &   19.0  &  3.60 & Thevenin \\
\hline
Mean:&&&&3.51$\pm$0.13&\\
\hline
Ba\,{\sc ii}&&&&&\\
6496.90 & 0.60 & $-$0.37  &   14.0  &  2.49 & Luck \\
\hline
\end{tabular}
\medskip
$^a$Normalized such that log$\Sigma \mu_{i} \epsilon(i)$ = 12.15\\
\medskip
$^b$Sources of $gf$-values\\
\medskip
$^c$Spectrum synthesis
\end{table}
\begin{table}
%\contcaption{}
References:
\medskip
\begin{tabular}{lll}
Jeffery && Jeffery, Woolf, \& Pollacco (2001)\\
Kurucz && Kurucz's database\\
Luck && Compilations by R. E. Luck \\
NIST&& NIST database\\
Thevenin && Th\'{e}venin (1989, 1990)\\
Wiese    && Wiese, Fuhr, \& Deters (1996)\\
\hline
\end{tabular}
\end{table}


\begin{thebibliography}{99}

\bibitem{} Asplund, M., Gustafsson, B., Lambert, D. L., Rao, N. K., 2000, A\&A, 353, 287
\bibitem{} Bond, H. E., Luck, R. E., Newman, M. J., 1979, ApJ, 233, 205 
\bibitem{} Drilling, J. S., 1973, ApJ, 179, L31
\bibitem{} Drilling, J. S., Sch\"{o}nberner, D., Heber, U., Lynas-Gray, A. E., 1984, ApJ, 278, 224
\bibitem{} Jeffery, C. S., Starling, R. L. C., Hill, P. W., Pollacco, D. L., 2001, MNRAS, 321, 111
\bibitem{} Jeffery, C. S., Woolf, V. M., Pollacco, D. L., 2001, A\&A, 376, 497
\bibitem{} Pandey, G., Rao, N. K., Lambert, D. L., Jeffery, C. S., Asplund, M., 2001, MNRAS, 324, 937
\bibitem{} Pandey, G., Lambert, D. L., Rao, N. K., Jeffery, C. S., 2004, ApJ, 602, L113
\bibitem{} Pandey, G., Lambert, D. L., Jeffery, C. S., Rao, N. K., 2006, ApJ, 638, 454
\bibitem{} Rao, N. K., Lambert, D. L., 2003, PASP, 115, 1304 
\bibitem{} Stephenson, C. B., Sanduleak, N., 1971, Publ. Warner \& Swasey Obs. 1, 1
\bibitem{} Th\'{e}venin, F., 1989, A\&AS, 77, 137
\bibitem{} Th\'{e}venin, F., 1990, A\&AS, 82, 179
\bibitem{} Vanture, A. D., Zucker, D., Wallerstein, G., 1999, ApJ, 514, 932 
\bibitem{} Wiese, W. L., Fuhr, J. R., Deters, T. M., 1996, {\it J. Phys. Chem. Ref. Data}, Monograph No. 7

\end{thebibliography}

\begin{thebibliography}{99}

\bibitem{} Allende Prieto, C., Lambert, D. L., Asplund, M., 2002, ApJ, 573, L137
\bibitem{} Asplund, M., Gustafsson, B., Kiselman, D., Eriksson, K., 1997, A\&A, 318, 521
\bibitem{} Asplund, M., Gustafsson, B., Lambert, D. L., Rao, N. K., 2000, A\&A, 353, 287
\bibitem{} Asplund, M., Ryde, N., 1996, in:
     Hydrogen deficient stars, Jeffery C.S., Heber U.
    (eds.). ASP conf. series vol. 96, p. 57
\bibitem{} Bauschlicher, C. W., Jr., Langhoff, S. R., Taylor, P. R., 1988, ApJ, 332, 531
\bibitem{} Bakker, E. J., Lambert, D. L., 1998, ApJ, 502, 417
\bibitem{} Carlsson, M., 1986, Uppsala Astronomical Report, No. 33
\bibitem{} Cayrel, R., 1963,  C. R. Acad. Sci., 257, 3309
\bibitem{} Cottrell, P. L., Lambert, D. L., 1982, ApJ, 261, 595
\bibitem{} Davis Sumner, P., Phillips John, G., 1963, The red system (A$^2$$\Pi$$-$X$^2$$\Sigma$) of the CN molecule,
     Berkeley Analyses of Molecular Spectra, Berkeley: University of California Press.
\bibitem{} Galavis, M. E., Mendoza, C., Zeippen, C. J., 1997, A\&AS, 123, 159
\bibitem{} Gustafsson, B., Asplund, M., 1996, in:
     Hydrogen deficient stars, Jeffery C.S., Heber U.
    (eds.). ASP conf. series vol. 96, p. 27
\bibitem{} Gustafsson, B., Bell, R. A., Eriksson, K., Nordlund, A., 1975, A\&A, 42, 407
\bibitem{} J{\o}rgensen, U. G., Larsson, M, 1990, A\&A, 238, 424
\bibitem{} Lambert, D. L., Rao, N. K., Giridhar, S., 1990, JAA, 11, 475
\bibitem{} Lambert, D. L., Rao, N. K., 1994, JAA, 15, 47
\bibitem{} Lawson, W. A., Cottrell, P. L., Clark, M., 1991, MNRAS, 251, 687
\bibitem{} Liu, X. W., Barlow, M. J., Danziger, I. J., Clegg, R. E. S., 1995, MNRAS, 273. 47
\bibitem{} Luck, R. E., Lambert, D. L., 1981, 245, 1018
\bibitem{} Mihalas, D. 1978, Stellar Atmospheres, Freeman \& Co.:San Francisco
\bibitem{} Moore, Ch. E.: 1972, A Multiplet Table of Astrophysical Interest,
    NSRDS - NBS 40, Washington
\bibitem{} Moore, Ch. E.: 1993, Tables of Spectra of Hydrogen, Carbon, Nitrogen, and
    Oxygen Atoms and Ions, Editor: Jean W. Gallagher, CRC Series in Evaluated Data
    in Atomic Physics, CRC press
\bibitem{} Nave, G., Johansson, S., Learner, R. C. M., Thorne, A. P., Brault, J. W., 1994, ApJS, 94, 221
\bibitem{} Rao, N.K., Lambert, D.L., 1990, AJ, 105, 1915
\bibitem{} Rao N. K., Lambert D. L.,1996, in:
    Hydrogen deficient stars, Jeffery C.S., Heber U.
    (eds.). ASP conf. series vol. 96, p. 43
\bibitem{} Rao N. K., Lambert D. L., 1997, MNRAS, 284, 489
\bibitem{} Rao N. K., Lambert D. L., Adams, M. T., Doss, D. R., Gonzalez, G., Hatzes, A. P.,
    James, R., Johns-Krull, C. M., Luck, R. E., Pandey, G., Reinsch, K.,
    Tomkin, J., 1999, MNRAS, 310, 717
\bibitem{} Sch\"{o}nberner, D., 1975, A\&A, 44, 383
\bibitem{} Searle, L., 1961, ApJ, 133, 531
\bibitem{} Swensson, J. W., Benedict, W. S., Delbouille, L., Roland, G., 1970, The
    Solar Spectrum from $\lambda$7498 to $\lambda$12016. A Table of Measures and
    Identifications. Mem. Soc. R. Sci. Li\`{e}ge, Spec. Vol. No. 5
\bibitem{} Tull, R. G., MacQueen, P. J., Sneden, C. and Lambert, D. L., 1995, PASP,
    107, 251
\bibitem{} Wiese, W. L., Fuhr, J. R., Deters, T. M., 1996, Journal of Physical and Chemical Reference Data, Monograph No. 7

\end{thebibliography}
\end{document}